\documentclass[12pt,preprint]{aastex}

\begin{document}

\title{Spatial Variations of Galaxy Number Counts in the SDSS I.:
Extinction, Large-Scale Structure and Photometric Homogeneity}

\author{
Masataka Fukugita\altaffilmark{1},
Naoki Yasuda\altaffilmark{2},
Jon Brinkmann\altaffilmark{3},
James E. Gunn\altaffilmark{4},
$\check{\rm Z}$eljko Ivezi\' c\altaffilmark{4}, 
Gillian R. Knapp\altaffilmark{4},
Robert Lupton\altaffilmark{4}, and
Donald P. Schneider\altaffilmark{5}
}

\altaffiltext{1}{Institute for Cosmic Ray Research, University of Tokyo, Kashiwa, 277 8582 Japan; fukugita@icrr.u-tokyo.ac.jp}
\altaffiltext{2}{National Astronomical Observatory, Mitaka, Tokyo, 181 8588, Japan}
\altaffiltext{3}{Apache Point Observatory, Sunspot, NM 88349, U. S. A.}
\altaffiltext{4}{Princeton University Observatory, Princeton, NJ 08544, U. S. A.}
\altaffiltext{5}{Department of Astronomy and Astrophysics, Pennsylvania 
State University, University Park, PA 16802, U. S. A.}

\begin{abstract}
We study the spatial variation of galaxy number counts using five
band photometric images from the Sloan
Digital Sky Survey. The spatial variation of this sample of
46 million galaxies collected from 2200 sq. degrees can be understood 
as the combination of Galactic extinction and
large-scale clustering. With the use of the reddening map of
 Schlegel, Finkbeiner \& Davis\markcite{SFD98} (1998), the
standard extinction law is verified for the colour bands from 
$u$ to $z$ within 5\% in the region of small extinction values, 
$E(B-V)<0.15$.
The residual spatial variations of the number counts suggests  
that the error of global calibration for SDSS photometry is  
smaller than 0.02 mag. 
\end{abstract}

\keywords{dust, extinction --- cosmology: large-scale structure of universe --- techniques: photometric}

\section{Introduction}

The Sloan Digital Sky Survey (SDSS; York et al.\markcite{York00} 2000)  
is conducting photometric and spectroscopic surveys over about 
$\pi$ steradians of the sky, producing a homogeneous 
data base of galaxies and other astronomical objects
with accurate astrometric calibrations (Pier et al.\markcite{Pier02} 2002). 
The data published as {\it Data Release 1} (DR1; Abazajian et al.\markcite{DR1paper} 2003;
see also http://www.sdss.org/dr1/index.html) comprise 2196 sq. deg.
of sky, which is about 25\% of the survey goal.
One of the most important features
of the SDSS is its homogeneous photometry by virtue of 
a large format mosaic CCD camera (Gunn et al.\markcite{Gunn98} 1998),
and real-time calibration of the photometry using a 50 cm
Photometric Telescope (Hogg et al.\markcite{Hogg01} 2001).

In this paper we study the spatial variation of galaxy number 
counts observed in the photometric survey of the SDSS. 
The causes of the spatial variation are (i) spatial variation
of extinction due to the dust in the Galaxy, (ii) large-scale
structure of the Universe, and (iii) errors of photometric
calibrations, especially those for stripe to stripe (or segments
of stripes), 
since the observations are carried out along great circle scans 
(York et al. 2000).
The prime motivation of the present work is to investigate 
(iii), but to acomplish this goal
it is necessary to remove the effects of (i) and (ii), 
both which have fundamental scientific significance.

The most modern map of Galactic reddening over the sky 
is that of Schlegel, Finkbeiner \& Davis\markcite{SFD98} (1998; SFD),
which was produced by merging the COBE/DIRBE and IRAS/ISSA FIR maps. 
The extinction curve is given by
$A_\lambda=k(\lambda) E(B-V)$, where $E(B-V)$ is the reddening.
An important aspect of our study is to examine whether $A_\lambda$
correctly describes extinction due to Galactic dust. 
The most model-independent test for the extinction correction can be made 
with galaxy number counts (e.g., Burstein \& Heiles\markcite{BH82} 1982).
For this test, it is desirable to work with galaxy number counts as 
faint as possible, so that the galaxy density is sufficiently high and
the distribution of galaxies is sufficiently smooth that large-scale
clustering effects are small. 
We choose the range of magnitude as  $r=18.5-20.5$ mag
(median redshift is 0.25), which
is reasonably faint, yet photometric measurements are made at a
high signal to noise ratio and the star-galaxy classification is 
hardly affected by the variation of seeing in 
the photometric survey in the SDSS. 

Even at this faint magnitude we expect the effect of large-scale 
clustering to be appreciable. The importance of the effect, however, 
can be predicted from the angular two-point correlation function 
(Connolly et al.\markcite{Connolly02} 2002).

The spatial variations in galaxy number counts that remain after subtraction 
of Galactic extinction and large-scale structure may be ascribed to
variations of the photometric calibration, which is made stripe by stripe,
or its segments when the scan of the stripe was obtained over
several nights. The challenge is to separate
these three effects, which we shall consider in this paper.

\section{Galactic extinction} 

We show in Table 1 the data used in the present analysis. Stripes \#9$-$12
are the northern equatorial stripe (\#10) and neighbouring stripes, \#82
is the southern equatorial stripe, \#76 and \#86 are stripes in the southern
sky, and the others are stripes in the northern hemisphere. The total area
is 2196 sq. deg. 
We use the data processed with the photometric pipeline version
v5.3, as given in DR1. We refer to Strauss et al.\markcite{Strauss02} (2002) for the selection 
of galaxies from the photometric catalogue.
The galaxy catalogue thus produced contains a small fraction of fake 
objects at bright magnitude $r'<16$, but the contamination becomes 
negligible for fainter magnitudes which are considered in this paper; 
see Yasuda et al.\markcite{Yasuda01} (2001). 

A survey stripe is 2.52 deg wide; scans can be up to 120 deg
in length, and the stripe consists of 
a pair of TDI scan operations 
[denoted by (N,S) in Table 1] to fill the gaps between CCD chips. 
The imaging is made with the SDSS
$ugriz$ filters (Fukugita et al.\markcite{Fukugita96} 1996).
We work primarily with the $r$ band data. We divide the stripe into 
square regions of 2.5 deg$^2$ along the
stripes, and count galaxies contained in these square regions. The number
of galaxies with $18.5\leq r\leq 20.5$ mag
contained in a square is approximately 11700;
therefore the Poisson noise is about 0.4\%
(the corresponding magnitude offset, $\Delta m\simeq 0.004$ mag, is 
negligible for this work). 

We first derive the differential number count $\overline{N}(m)$
from the entire sample employing the reddening map of SFD and the
default standard extinction
law (explained below). 
We then calculate the number count for each $2.5^\circ$ square region
without applying the extinction correction, and
using the reference count curve $\overline{N}(m+\Delta m)$ with magnitude
offset $\Delta m$, we fit it to the four data points in the range 
$r=18.5-20.5$ mag devided into 0.5 mag bins by adjusting $\Delta m$ with
a chi-squares fit.

We show in Figure 1 $\Delta m$ 
versus Galactic extinction, 
$A_r^{\rm SFD}$,  calculated
from the reddening map of SFD assuming $k(r)=A_r/E(B-V)=2.75$
from the standard extinction curve (SFD; O'Donnell 1994\markcite{ODonnell94};
see also Cardelli, Clayton \& Mathis 1989\markcite{CCM89}; Fitzpatrick 1999\markcite{Fitz99}) 
and calculating the mean
of extinction over the $(2.5^\circ)^2$ region.
The individual data points and their binned mean in $A_r^{\rm SFD}$
are plotted. The error bars show the rms scatter. Although the scatter
is significant, there is a clear trend that the mean of $\Delta m$ and
$A_r$ are nearly identical, showing that 
the observed variation of galaxy counts 
is primarily due to Galactic extinction.
Hence we may identify $\langle \Delta m\rangle=A_r^{\rm counts}$.
Unfortunately, the scatter is so large compared to the fitting range of
the abscissa 
that it is not possible to perform a two-parameter fit. So we first
fit the data (using all data points) to determine the slope,
enforcing that the curve passes
through the origin (as the plot indicates), and then vary the 
constant while the slope is fixed. 
The fit yields 
$k(r')^{\rm counts}=A_r^{\rm counts}/E(B-V)=2.64\pm 0.11$
and the zero point that gives $\chi^2$ minimum is 0.007$\pm$0.005.
The constant term is sufficiently small that it can be ignored. The 
measured slope verifies the $k(r)$ from the standard extinction law 
together with the reddening map of SFD to an accuracy of 5\%. 
The rms scatter around the line is 0.096 mag.

A similar analysis is made for the $u$, $g$, $i$ and $z$ colour bands. We 
selected the magnitude ranges of 18.5$-$20.5 for $u$ and $g$, 
18.0$-$20.0 for $i$ and 
17.5$-$19.5 for $z$. The resulting plots are presented in Figure 2.
The slope of the $\chi^2$ fits and the dispersion around the curves 
as normalised by $k(\lambda)$ are summarised in Table 2.
For all colour bands  $k(\lambda)^{\rm counts}$ does not deviate from 
$k(\lambda)$ by more than $\approx$5\%.
Note, however, that our test is limited to the sky with small reddening
of $E(B-V)<0.15$ (the SDSS is designed to avoid high extinction fields;
see York et al. 2000).

We may use $k(\lambda)^{\rm counts}$ to derive a constraint on the
$R=A_V/E(B-V)$ parameter that represents total-to-selective 
extinction. 
The extinction curves of Cardelli et al.\markcite{CCM89} (1989) and 
their updates given by 
O'Donnell\markcite{ODonnell94} (1994) contain the parameter $R$ [$R=k(V)$]
which is related to the greyness of the dust. Using O'Donnell's extinction 
curve, which is given by the form $A_\lambda/A_V=a_\lambda+b_\lambda/R$
with $a_\lambda$ and $b_\lambda$ functions of $\lambda$, we can derive
constraint on $R$, for example $R=2.98\pm0.23$ from the $r$-band.
The constraints on $R$ from other colour bands are given in Table 2.
All constraints are consistent with $R=2.8$ to 3.2.

Before we discuss the issue of the scatter
around the relation, we examine the reddening in the colour of  
galaxies for a consistency check.
Figure 3 shows $g-r$ colour of galaxies at the fifth percentile 
from the reddest
as a function of $E(B-V)$ predicted by SFD. The trend of reddening
against $E(B-V)$ is evident. The binned data points agree with the
line $g-r=1.04E(B-V)+$const., which is expected from the standard reddening
law.

A further consistency test is given in Figure 4, where 
$\Delta m_g-\Delta m_r$ 
($\langle\Delta m_g-\Delta m_r\rangle=(A_g-A_r)^{\rm count}$),
which is derived from
the extinction detected in the number count data, is plotted against
$E(g-r)^{\rm galaxy~colour}=
(g-r)_{\rm red~5\%}-(g-r)_{\rm red~5\%}|_{E(B-V)=0}$ calculated
from Fugure 3. The plot is consistent
with $(A_g-A_r)^{\rm count}=E(g-r)^{\rm galaxy~colour}$ (indicated by
the dotted line), although the scatter is large and we see some
data points that are scattered into the negative regions. 
Note that this figure shows a consistency of reddeing and extinction
derived solely using the galaxy data, independent from the external
data of $E(B-V)$ from Galactic far infra-red emission.

\section{Spatial variations from large-scale structure}

The rms scatter observed in the previous section is 
larger, by an order of magnitude, than that expected 
from the Poisson noise.
To investigate
whether the scatter is due to large-scale clustering of galaxies, 
we study the variation of galaxy counts in the $r$ band 
varying area sizes of sky regions.

We define circles with radii $r=1.25^\circ$ along the stripes,
and decrease the radius keeping the centres of the circles fixed\footnote{We
prefer to use circles rather than squares for the simplicity to interpret
the data in terms of angular correlation functions.}.
We count 
numbers of galaxies with $r=18.5-20.5$ grouped into four 0.5 mag bins
contained in those circular areas
after applying an extinction correction according to the SFD map 
[assuming $k(r)=2.75$], 
and then calculate the corresponding magnitude offset
$(\Delta m)_c$ in the same way as above.

The rms scatter, $(\Delta m)_c$, thus obtained is plotted as 
a function the area $\Omega=\pi r^2$ in Figure 5. 
The error bars show the Poisson statistics.
The rightmost point (open square) refers to the scatter observed
in 2.5 deg square (6.25 sq. deg) which was seen in the previous section.
We also added one more point for 3 sq. deg (open square)
for comparison. (The data for square areas
give values about $\approx$3\% smaller than those for circular regions having
the same area, as expected from the angular correlation function; see below.)
The size of $(\Delta m)_c$ is larger than the Poisson noise
at least by an order of magnitude.
To study whether $(\Delta m)_c$, which increases as the area size
decreases, can be attributed to the 
effect of large-scale clustering, 
we calculate the expectation from the empirical
angular two-point correlation function $w(\theta)$. The
fluctuations of the number of galaxies observed in area $\Omega$ are given by
\begin{equation}
\langle (N-\nu\Omega)^2\rangle=\nu\Omega+\nu^2\int_\Omega d\Omega_1 d\Omega_2
                    w(\theta_{12}),
\label{eq:1}
\end{equation}
where $\nu$ is the number of galaxies per unit solid angle.
The rms of the count 
$\delta N/\overline{N}=\langle(N-\nu\Omega)^2\rangle^{1/2}/\nu\Omega$
is translated to $\Delta m$ using 
$\Delta m=\left(dN/dm\right)^{-1}\Delta N=(\alpha\ln10)^{-1}\Delta N/N$
where $\alpha\simeq 0.42$ is the empirical slope of the count, 
$N\sim 10^{\alpha m}$, at the relevant magnitude range.
%

The angular correlation function estimated by 
Connolly et al.\markcite{Connolly02} (2002)
from the analysis for the northern equatorial stripe of the SDSS is
\begin{equation}
w(\theta)=A_w \theta^{1-\gamma},
\label{eq:4}
\end{equation}
where $1-\gamma=-0.722\pm0.031$ and $\log A_w=-2.13\pm0.13$ for 
$r^*=19-20$ mag with the preliminary photometric calibration 
(Smith et al.\markcite{Smith02} 2002; Stoughton\markcite{Stoughton02} et al. 2002). 
This angular correlation function is 
consistent,  within the errors, with earlier 
analyses (Stevenson et al.\markcite{Stevenson85} 1985; Maddox, Efstathiou
\& Sutherland\markcite{Maddox93} 1990; 
Couch, Jurcevic \& Boyle\markcite{Couch93} 1993). 
Connolly et al. could show the power law only for $\theta<1^\circ$. Maddox
et al., showed that the correlation function shows a break at around
2-2.5$^\circ$ from thier APM data. 
We, therefore, model the angular correlation function
by a two-power law that breaks at 2$^\circ$, with the second power slope
$1-\gamma\approx -2.1$ consistent with the APM result.

Figure 5 shows that the rms scatter of galaxy number counts
in different parts of the sky after correcting for Galactic extinction is
consistent with that which is expected from the angular two-point
correlation function integrated over circular areas. 
This means that a significant scatter 
around the extinction fits of Figures 1 and 2 is produced by large-scale 
clustering of galaxies.

\section{Homogeneity of the SDSS photometric calibration}

We now investigate the residual spatial variation of galaxy
number counts integrated over segments of the stripes (some of the runs
are merged) given in Table 1.    
Figure 6 shows the offset after the extinction correction
$(\Delta m)_c$ calculated from the galaxy counts for specific segments
of the stripes relative to the reference count. 
The error bars represent the standard deviation expected from the
two-power model of the angular correlation function with a cut-off
at $\theta=5^\circ$ in the integral (\ref{eq:1})
over the rectangular region of the segment.

The figure shows that the mean still scatters by $\sim\pm0.04$ mag, but
the majority of the data are consistent with zero if we consider 
the variation expected from large-scale clustering, 
i.e., 19 out of 32 data points are within 1 $\sigma$, 
and the maximum deviation
is 1.6 $\sigma$. This implies a null detection of photometric 
calibration errors. We may make this statement more quantitative
by applying the statistics in the following way. 
We estimate the likelihood that the observed $(\Delta m)_c$ distribution 
is consistent with the Gaussian distribution of the dispersion
$\sigma_{\rm eff}=\sqrt{\sigma^2+(\delta m)^2}$, where $\delta m$ represents
the error, other than that from large-scale structure, 
that increases the dispersion. We then calculate
the probability as a function of $\sigma_{\rm eff}$, and find that the
probablity of the observed distribution being consistent with 
the Gaussian at 68\% confidence level (1 $\sigma$) only when
$\delta m<0.019$ mag.
This is taken as the upper limit on the error from photometry.
This error meets the design requirement of the SDSS photometry; the
global variation of the calibration error is no more than random
errors of photometry. 

We do not carry out similar analyses for other colour bands, since
the evaluation of the angular two-point correlation is only
available in the $r$ bands. Table 2 displays the  
continuous increase of the rms scatter from the $z$ to the $u$ band.
Since it is known that the photometric accuracy of the
$g$, $r$, and $i$ bands
are comparable (DR1), we would ascribe this increase to 
increasingly stronger
large-scale correlation in bluer bands. For the $u$ band, random
photometric errors ($\sim 0.04$ mag) may also contribute to the rms
scatter, but they are smaller than the increase of the rms scatter we observed
from the $g$ to the $u$ band.

In conclusion, we have demonstrated that the spatial variation of galaxy
number counts are understood as the sum of Galactic extinction and
large-scale clustering of galaxies. The analysis verified the validity
of the SFD extinction map and the standard extinction 
law from $u$ to $z$ band within 5\% in regions of small
reddening $E(B-V)<0.15$. We find that the $R$ parameter lies in
the range 2.8$-$3.2 for the extinction curve of O'Donnell\markcite{ODonnell94} (1994).
Finally we do not detect systematic errors in the global
SDSS photometric calibration at least those in excess of 0.02 mag
from run to run. 

\acknowledgments

Funding for the creation and distribution of the SDSS Archive has been provided
by the Alfred P. Sloan Foundation, the Participating Institutions, the National
Aeronautics and Space Administration, the National Science Foundation, the U.S.
Department of Energy, the Japanese Monbukagakusho, and the Max Planck
Society. The SDSS Web site is http://www.sdss.org/. 
The SDSS is managed by the Astrophysical Research Consortium (ARC) for the
Participating Institutions. The Participating Institutions are The University of
Chicago, Fermilab, the Institute for Advanced Study, the Japan Participation
Group, The Johns Hopkins University, Los Alamos National Laboratory, the
Max-Planck-Institute f\"ur Astronomie, the Max-Planck-Institut f\"ur
Astrophysik, New Mexico State University, University of Pittsburgh,
Princeton University, the United States Naval Observatory, and 
the University of Washington.
MF is supported in part by the Grant in Aid of the
Japanese Ministry of Education.



\clearpage

\begin{table}
\caption{Stripes used for this analysis.}
\begin{tabular}{lll}
\tableline
stripe & segments  & run: (N,S) pairs\\
\tableline
9 & 1 & (1140,1231)\\
10 (N. eq. stripe) & 3 & (756,1239), (756+745, 752),  (745+752)\\
11 & 2 & (1907,1462), (1992+1458,1462)\\
12 & 4 & (2126,2125+2247), (2126,2247)\\
   &   & (2190,2247),(2190,1478)\\
\tableline
34 & 1 & (2137, 2131+2243)\\
35 & 3 & (2076, 1895)\\ 
   &   & (2076+1895, 1889+2075+2074+1896)\\
   &   & (2299+2305, 2326+2328)\\
36 & 5 & (1345,1331), (1345,1332) \\
   &   & (2189+1345,2078+2134)\\
   &   & (1345,2238), (2335,2248)\\
37 & 4 & (1402+1450,1350), (1412,1350)\\
   &   & (1412,2206), (1453,2207)\\
42 & 1 & (1336,1339)\\
43 & 1 & (1356,1359)\\
\tableline
76 & 1 & (1043,1035)\\
82 (S. eq. stripe) & 1 & (2738+2662,3325)\\
86 & 5 & (1659,1737), (1659+1891+1740,1729)\\
   &   & (1741,1729), (1893+1869,1729)\\
   &   & (1869+1045,1729)\\
\tableline
\end{tabular}
\end{table}

\clearpage

\begin{table}
\caption{The extinction functions obtained from the galaxy
number counts and those from the standard extinction curve.}
\begin{tabular}{lrrrrr}
\tableline
colour bands  & $u~~$  & $g~~$  & $r~~$ & $i~~$ & $z~~$ \\
\tableline
standard extinction curve $k(\lambda)$ 
                                    & 5.155 & 3.793 & 2.751 & 2.086 & 1.479 \\
$k(\lambda)^{\rm counts}/k(\lambda)$& 1.021 & 0.995 & 0.959 & 0.938 & 0.952 \\
                & $\pm 0.036$ & $\pm 0.041$ & $\pm 0.040$ & $\pm 0.048$ & $\pm 0.060$ \\
 rms scatter                        & 0.143 & 0.125 & 0.091 & 0.083 & 0.073 \\
constraint on $R$ &  3.21 & 3.08 & 2.98 & 2.94 & 2.99 \\
                  & $\pm0.39$ & $\pm0.30$ & $\pm0.23$ &$\pm0.24$ &$\pm0.27$\\
\tableline
\end{tabular}
\end{table}

\clearpage


\begin{figure}
\plotone{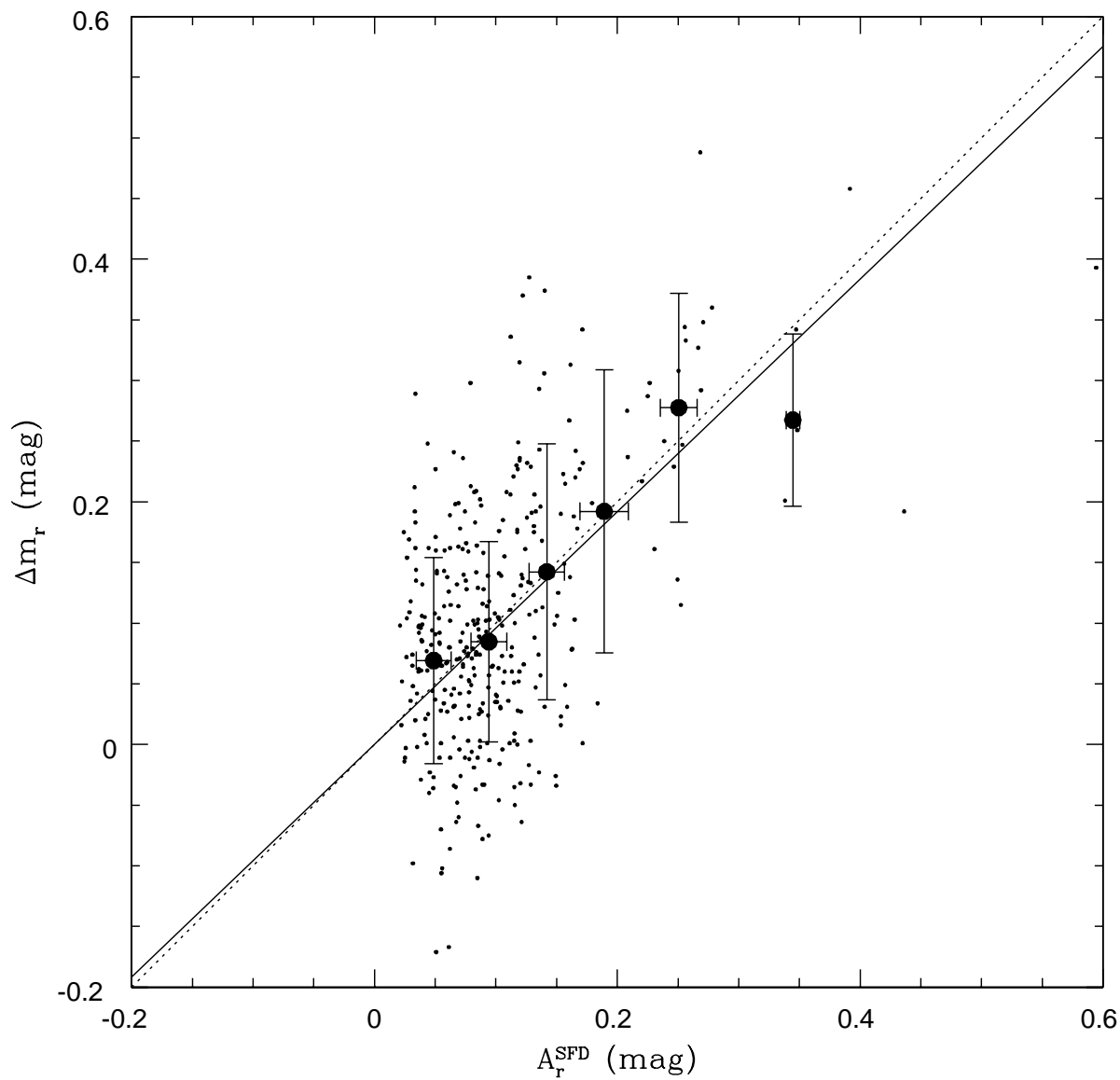}
\caption{
Magnitude offsets corresponding to the variation of galaxy number
counts in 2.5 degree square fields plotted against mean 
extinction calculated
from the SFD reddening map and the standard extinction 
curve $A_r^{\rm SFD}$.
Larger circles are
the mean in bins of $A_r^{\rm SFD}$, and the error bars show
the rms. The solid line is the result
of a $\chi^2$ fit, and
dotted line is the identical regression line 
$\langle \Delta m\rangle\equiv A_r^{\rm counts}=A_r^{\rm SFD}$.
}
\end{figure}

\begin{figure}
\plotone{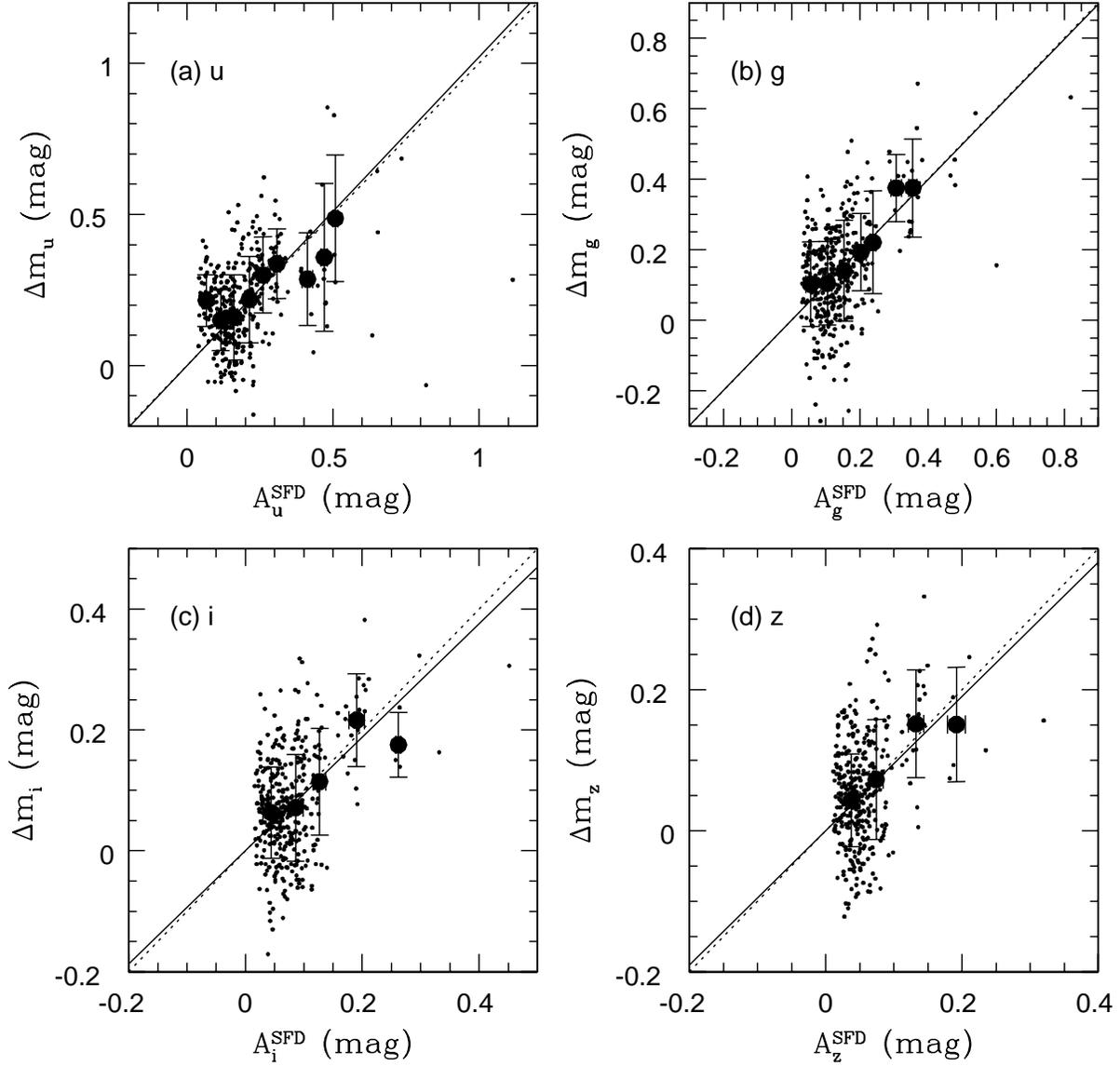}
\caption{
Same as Fig. 1, but for other four SDSS colour bands: (a) $u$ band,
(b) $g$ band, (c) $i$ band, and (d) $z$ band.
}
\end{figure}

\begin{figure}
\plotone{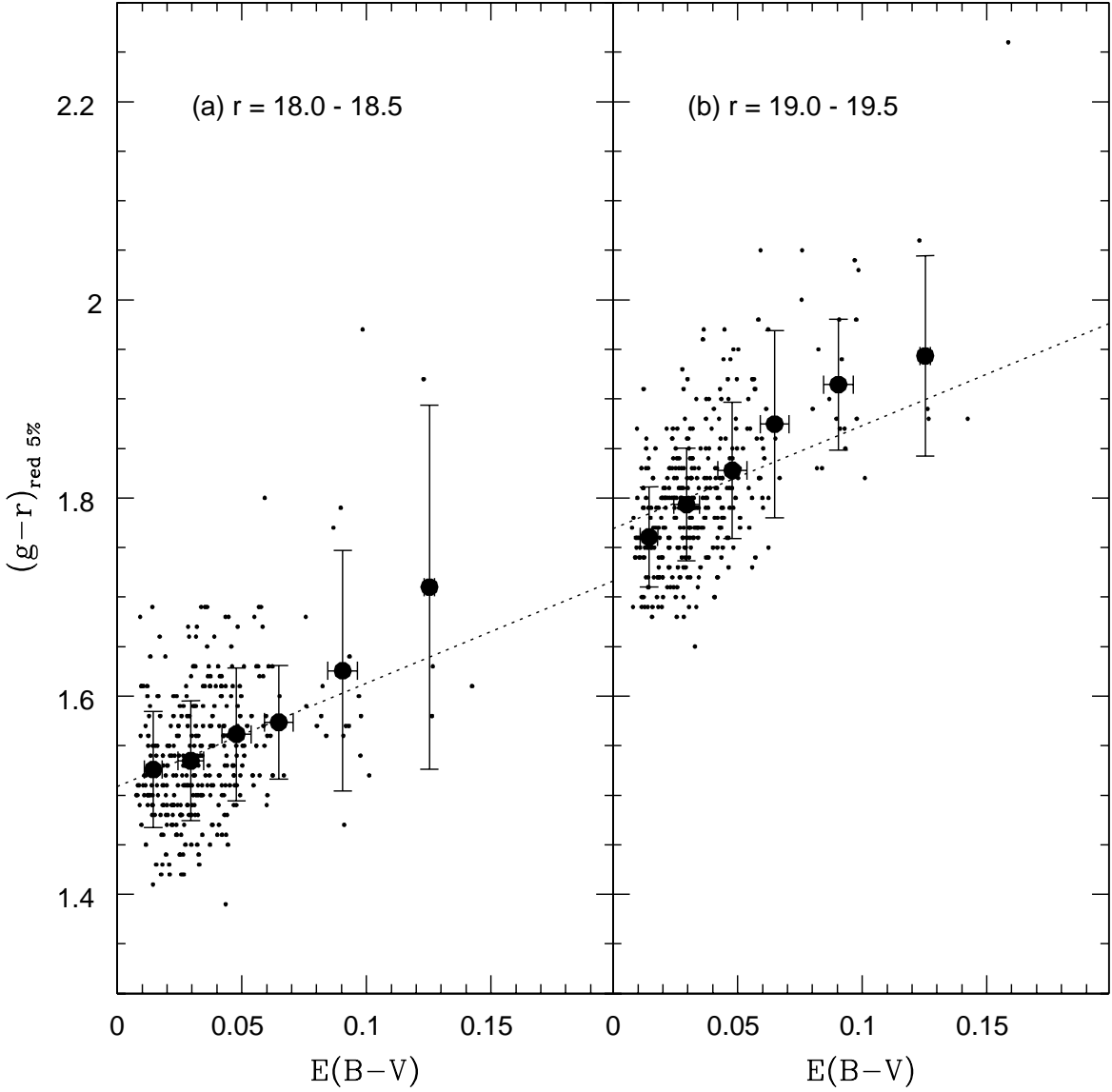}
\caption{
$g-r$ colours of galaxies at the 5$^{\rm th}$ percentile from the reddest
in 2.5 degree square regions 
plotted as a function of mean reddening obtained from the
map of SFD.  The figures are shown for two sets of galaxies
(a) $r=18-18.5$, and
(b) $r=19-19.5$. The lines represent $g-r=1.04E(B-V)+$const
expected for the standard reddening law.
}
\end{figure}

\begin{figure}
\plotone{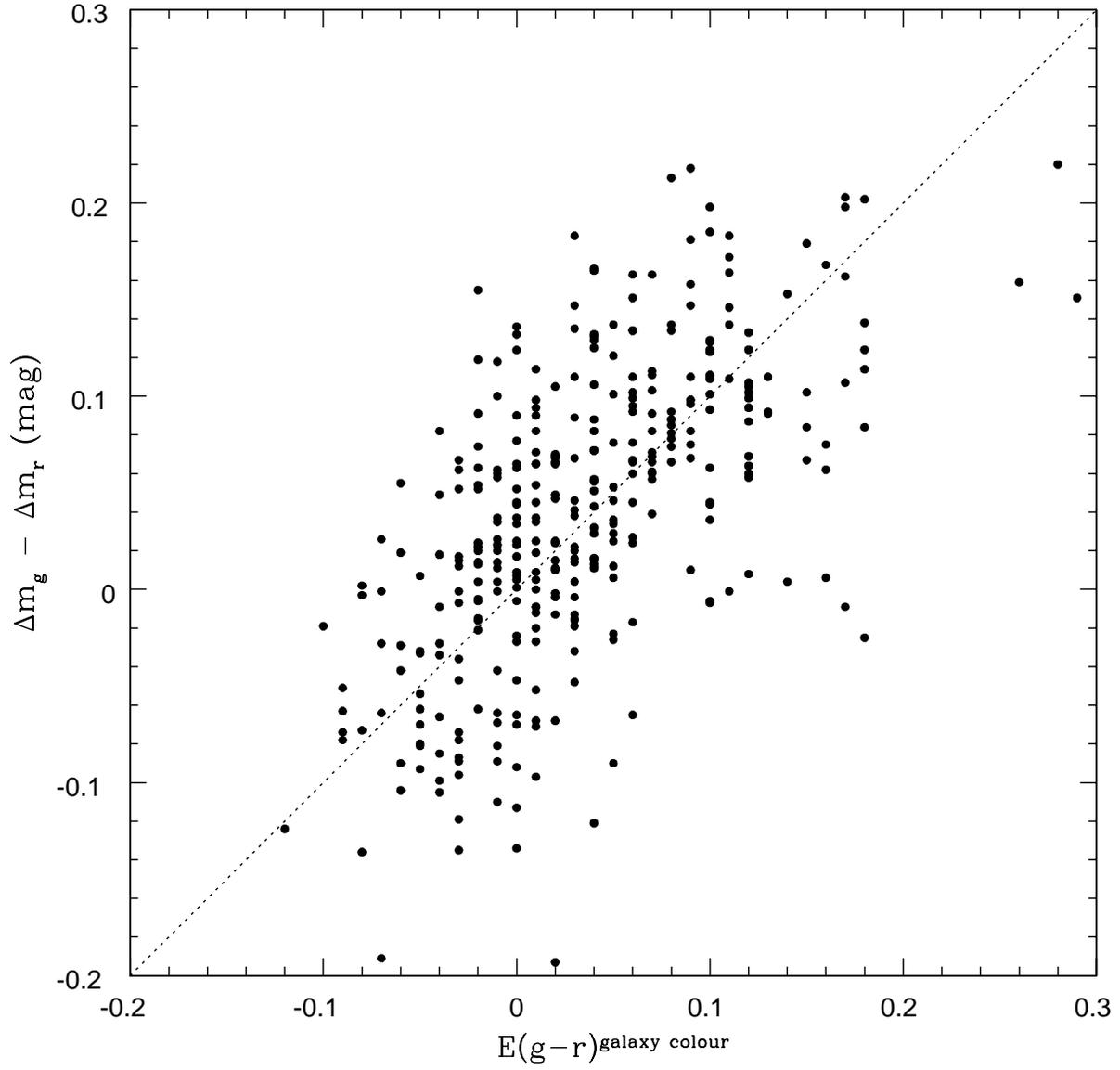}
\caption{The difference of the offset of the number counts
$\Delta m_g-\Delta m_r$, the mean of which is to be indetified with
$(A_g-A_r)^{\rm count}$, plotted against $E(g-r)^{\rm galaxy~colour}$
derived from the colour excess of galaxies. The dotted line 
shows the identical regression line 
$ \Delta m_g-\Delta m_r=E(g-r)^{\rm galaxy~colour}$.
}
\end{figure}

\begin{figure}
\plotone{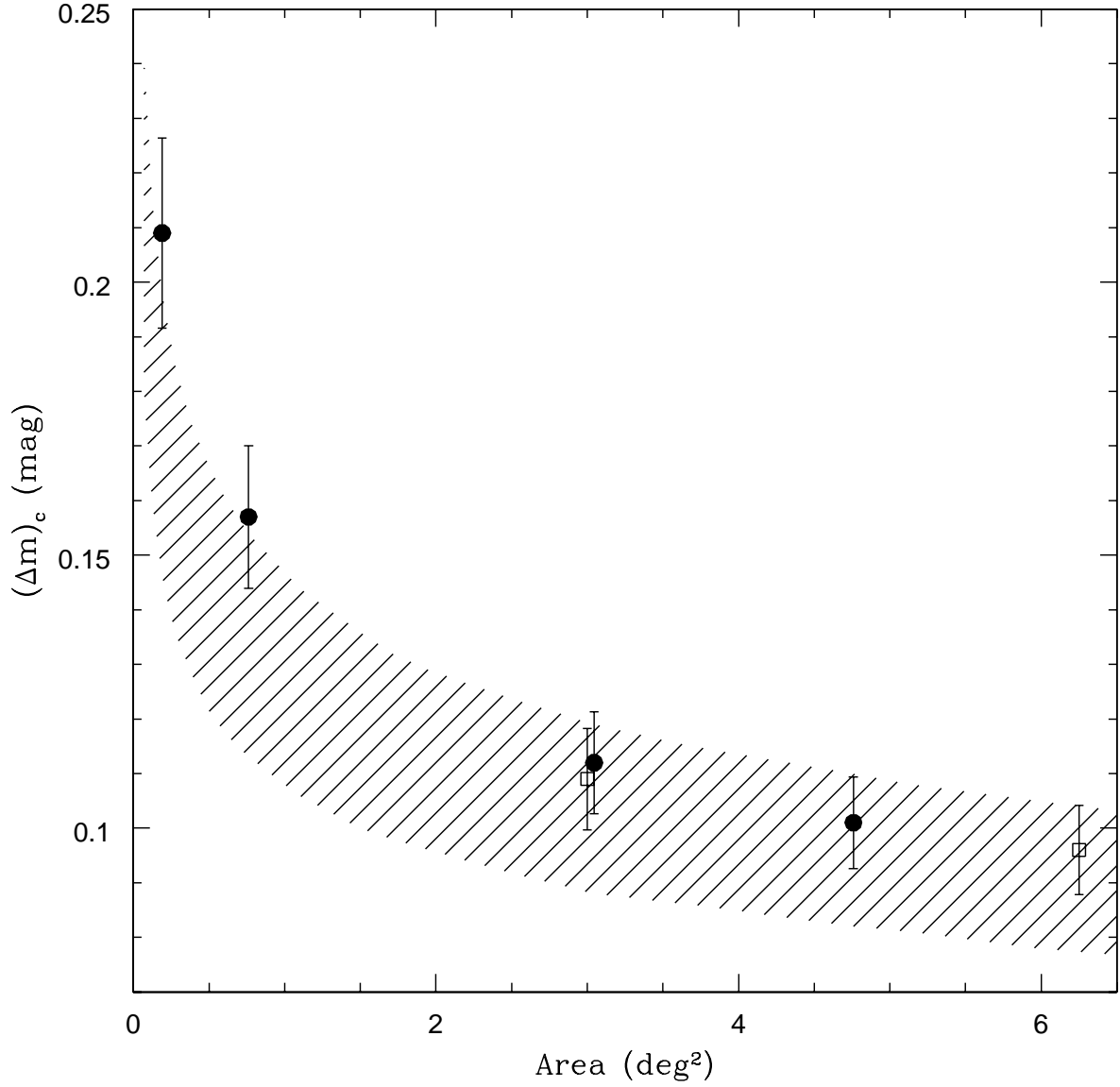}
\caption{
The excess of galaxy number counts (after applying the
extinction correction) as a function of the area
$\Omega r^2$ represented as the offset of magnitude from the
mean. The expectation from the angular two-point correlation
function is represented by shading.
}
\end{figure}

\begin{figure}
\plotone{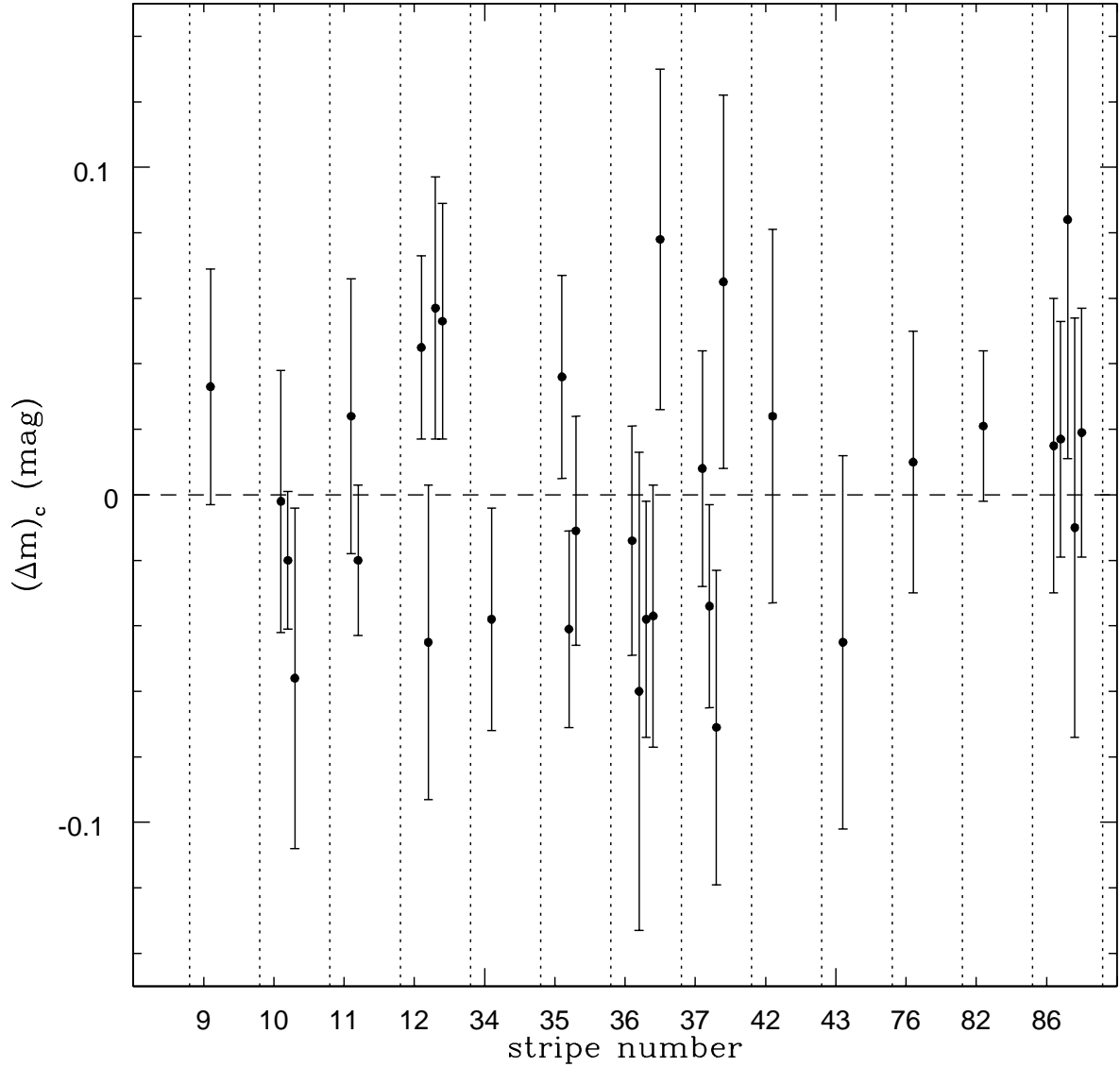}
\caption{
Magnitude offsets corresponding to the variation of galaxy 
number counts (after applying the
extinction correction) integrated over the
segment of stripes as defined in Table 1. The error bars
represent the variation expected from large-scale clustering
of galaxies.  The abscissa is the stripe number.
}
\end{figure}

\end{document}